\documentclass{WileyMSP-template}
\usepackage{pifont}
\usepackage{amsmath}
\usepackage{amssymb}
\usepackage{xcolor}
\begin{document}

\justifying
\setlength{\parindent}{0pt}

\makeatletter
\def\@cite#1#2{\textsuperscript{[{#1\if@tempswa , #2\fi}]}}
\makeatother

\pagestyle{fancy}
\rhead{\includegraphics[width=2.5cm]{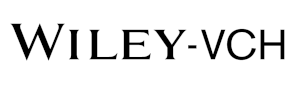}}

\title{Simultaneous Generation of Quantum Frequency Combs across Distinct Modal Families in a Single Silicon Nitride Whispering Gallery Mode Resonator}

\maketitle



\author{Bo Ji}
\author{Nianqin Li}
\author{Yongjun Yang}
\author{Tengfei Wu*}
\author{Guangqiang He*}



\begin{affiliations}

Bo Ji, Nianqin Li, Prof. Guangqiang He\\
State Key Laboratory of Advanced Optical Communication Systems and Networks, Department of Electronic Engineering, 
Shanghai Jiao Tong University, Shanghai 200240, China\\
Email Address: gqhe@sjtu.edu.cn \\
Yongjun Yang, Tengfei Wu\\
Science and Technology on Metrology and Calibration Laboratory, Changcheng Institute of Metrology Measurement, Aviation Industry Corporation of China, Beijing 100095, China\\
Email Address: tengfei.wu@163.com \\
Prof. Guangqiang He\\
State Key Laboratory of Precision Spectroscopy, East China Normal University, Shanghai 200062, China


\end{affiliations}



\begin{abstract}


Quantum frequency combs (QFCs) are versatile resources for multi-mode entanglement, such as cluster states, crucial for quantum communication and computation. On-chip whispering gallery mode resonators (WGMRs) can generate these states at ultra-low threshold power. This work demonstrates the simultaneous generation of multiple QFCs using a single on-chip silicon nitride WGMR across distinct modal families. It presents a micro-ring resonator with a radius of 240 $\mathrm{\mu m}$, capable of supporting four modal families within the 130 to 260 $\mathrm{THz}$ frequency range for consistency regulation. The results indicate that, by carefully designing the structure of silicon nitride WGMRs, it is possible to generate quantum entangled frequency combs across distinct modal families simultaneously using monochromatic pump light. It is achieved by modulating the pump mode profiles with a spatial light modulator (SLM) or an on-chip inverse-designed mode converter. This approach offers a simple and low-cost method to achieve higher-density entanglement integration on-chip.

\end{abstract}


\section{Introduction}
Optical quantum states featuring multimode entanglement within QFCs are pivotal in advancing quantum information science. Quantum Kerr optical frequency combs have demonstrated significant potential in quantum communication and computation, providing important tools for advancing these fields.\cite{kues2017chip,reimer2016generation,menicucci2008} While their contribution to a deeper understanding of quantum mechanics is still an active area of research, recent works have highlighted their promising applications in generating time-bin entangled circuits,\cite{xiong2015compact,reimer2016a} photon pair sources,\cite{samara2019high,yin2021frequency} quadrature-squeezed vacuum states,\cite{vaidya2020broadband,yang2021squeezed,wilken2024broadband,wen_five-partite_2015,gu_generation_2012} heralded single photons,\cite{wu2021integrated} graph states,\cite{ pysher2011,medeirosdearaujo2014,chen2014} and multi-user quantum networks.\cite{wen2022realizing,roslund2014}

The evolution from discrete, bulky optical setups to on-chip QFCs underscores a monumental shift, courtesy of advancements in integrated photonics. These on-chip systems not only overcome the maneuverability, integration, and scalability challenges posed by traditional free-space OPOs but also leverage the enhanced mode confinement and nonlinearity inherent in integrated micro-resonators. Recent studies focusing on the high degree of squeezing and exploration of quantum processes in soliton microcombs within nanophotonic devices have highlighted the substantial potential of these platforms.\cite{zhang2021squeezed,guidry2022quantum}

Silicon nitride ($\mathrm{Si_3N_4}$) stands out for its compatibility with CMOS technology, offering low loss and a broad transparent window, key attributes for the deployment in quantum applications.\cite{xiang2022silicon} However, the quantum potential of silicon nitride WGMRs has been somewhat underexplored, particularly in terms of structural modeling and simulation.\cite{xuan2016high,huang2019thermorefractive,vernon2015spontaneous,arbabi2013measurements} Integration with advanced CMOS technology facilitates the precise design of WGMR parameters such as coupling, loss rate, and dispersion through cavity structure engineering. Moreover, incorporating a thermoelectric heater during the CMOS fabrication process allows for fine-tuning each QFC mode by adjusting the WGMR structure and ambient temperature.\cite{xue2016thermal,joshi2016thermally,lnq2024} The bipartite entanglement criterion provides a method to quantitatively measure the degree of entanglement, offering valuable insights into how structural modifications and temperature adjustments can influence the entanglement performance of QFCs and can work as efficient feedback for structure redesign. This approach not only enriches the understanding of entanglement within integrated photonic systems but also paves the way for optimized quantum communication and computing technologies.\cite{xiang2021high} 

Traditional methods use pump lasers to pump fundamental modes ($\mathrm{TE_{00} / TM_{00}}$) of on-chip waveguides.\cite{ji2017ultra} However, WGMRs function as crystal systems, featuring what we call modal families. Each modal family can be seen as independent of one another, with no interference between them, assuming no cross modal families interactions. Modal families provide an excellent multiplex channel to support parallel quantum information, akin to Space Division Multiplexing (SDM) in multi-core optical fiber communication, recognized as the most efficient method to broaden capabilities.

In our work, we utilize SDM technology in the generation of quantum entangled frequency combs, enabling high-density entanglement generation on a single integrated $\mathrm{Si_3 N_4}$ WGMR. In our method, dispersion and coupling engineering of WGMRs are essential to realize multiple modal families and high entanglement dimensions. Dispersion engineering involves both structural and material dispersion. The on-chip WGMR, due to its easy fabrication properties, makes structural dispersion modulation highly convenient. We designed a WGMR waveguide structure with a cross-section that is approximately rectangular trapezoidal, with the add-through waveguide also having the same shape to support similar modal families. We implement the OPO theory to establish the quantum dynamics of third-order nonlinear WGMRs, including self-phase modulation (SPM), cross-phase modulation (XPM), and four-wave mixing (FWM).\cite{breunig2016three,guidry2023multimode,gouzien2020morphing} This allows for the extraction of resonator structure parameters and supports thermal adjustment. We successfully designed an on-chip WGMR with a radius of 240 $\mathrm{\mu m}$ that can support four modal families, three of them can be used as multiplexing channels. By tuning the pump laser and using a SLM, we can generate three bipartite entangled frequency combs with up to twelve channels (six pairs) in each modal family, ideal for multi-channel quantum information networks. Furthermore, we can map the entanglement distribution in each mode according to the bipartite entanglement criterion, motivating redesigns of the WGMR structure. The simulation of entanglement distribution control may pave a new avenue for the field of QFC design.

This article is organized as follows: Section 2 outlines the simulation model for $\mathrm{Si_3N_4}$ WGMR and introduces multi-modal family structure dispersion and WGMR-waveguide coupling engineering. Section 3 presents a theoretical model for third-order nonlinearity dynamics in the WGMR, including self-phase modulation (SPM), cross-phase modulation (XPM), and four-wave mixing (FWM), which will make further adjustments to the resonance formed by structural dispersion. The combination of these effects, called phase-matching conditions, will determine if quantum frequency combs can be generated. In Section 4, we discuss the quantum entanglement witness, clarifying whether entanglement exists in the generated quantum frequency combs. Our calculations and simulation results are presented in Section 5. Finally, in Section 6 and 7, we discuss and conclude our paper.

\section{$\mathbf{\chi^{(3)}}$ Microring Cavity Design}
\begin{figure}
  \includegraphics[width=\linewidth]{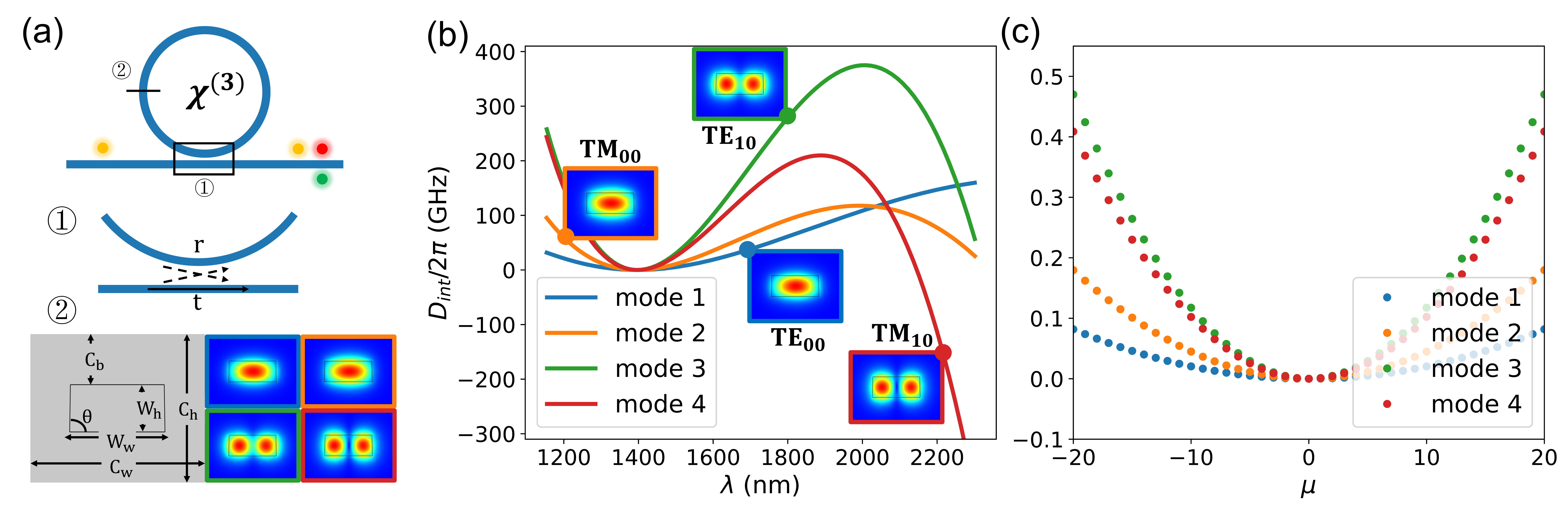}
  \caption{\label{fig:C0}(a) On-chip microring resonator coupling to add-through waveguide. The pump light is transmitted through the waveguide, partially coupled into the micro-ring cavity, and generates QFCs through the third nonlinearity, which are then coupled out. \ding{192} and \ding{193} respectively show the cross-sectional structures of the micro cavity in different regions. In \ding{192}, the structure of the coupling region between the micro cavity and the waveguide is depicted. This structure acts as a beamsplitter, where the transmittance and reflective coefficients, denoted as ${t}$ and ${r}$ respectively, remain flat due to the proximity of the pump, signal, and idler light frequencies considered in this paper. \ding{193} displays the cross-section of the micro-ring cavity, which features an $\mathrm{Si_3 N_4}$ waveguide encapsulated by an $\mathrm{Si O_2}$ cladding. The waveguide is structured in a trapezoidal shape, and its parameters are listed in Table~\ref{tab:tableC1}. The microring cavity we designed can support four modal families, including $\mathrm{TE_{00}}$, $\mathrm{TM_{00}}$, $\mathrm{TE_{10}}$, and $\mathrm{TM_{10}}$ modes. (b, c) Dispersion relationship of the four modal families ${D_{int}}$. The abscissa represent frequency and resonance peak distance reference to that one around $\mathrm{214.6~THz}$.}
\end{figure}

Figure~\ref{fig:C0}~(a) shows the general structure of the microcavity we designed. The on-chip microring cavity system consists of a ring waveguide and a straight waveguide.
The third-order nonlinearity of the waveguides material induces a four-wave mixing effect in the input pump light ($\Omega_{p1},\text{ }\Omega_{p2}$), resulting in the creation of signal ($\Omega_s$) and idler ($\Omega_i$) lights, with energy redistributed among different state modes. These processes must follow the laws of conservation of energy and momentum:
\begin{equation}
\begin{aligned}
 \Omega_{p1}+\Omega_{p2} &= \Omega_s+\Omega_i\\
 k_{p1}+k_{p2} &= k_s+k_i
\end{aligned}.
\end{equation}

 Figure~\ref{fig:C0}~(a) shows a standard structure of a microring resonator on chip featuring an add-through coupling configuration, relevant to the OPO theory. Our design uses $\mathrm{Si_3N_4}$ as the third non-linearity material, with the $\mathrm{Si_3N_4}$ waveguides embedded in a $\mathrm{SiO_2}$ cladding. This cladding layer acts as a protective cover and allows the integration of a micro-heater, closely attached to the resonator. The micro-heater facilitates the tuning and stabilization of resonance.\cite{xue2016thermal,joshi2016thermally} This structure is not difficult to fabricate in today's semiconductor industry.  

 Figure~\ref{fig:C0}~(a)~\ding{193} shows the cross-sectional structure of the microring. The cross-section of the ring is trapezoidal, defined by the height \({W_h}\), the width of the long side \({W_w}\), and the angle of the inclined side \({\theta}\). The trapezoidal encapsulated structure is filled with \(\mathrm{Si_3N_4}\) material, while the large outer area is filled with \(\mathrm{SiO_2}\) material. The parameters of the filled region are mainly controlled by three parameters: \({C_h}\), \({C_b}\), and \({C_w}\). The parameter \({C_b}\) represents the distance from the top plane of the \(\mathrm{Si_3N_4}\) waveguide to the top surface of the \(\mathrm{SiO_2}\) cladding, \({C_h}\) represents the thickness of the entire \(\mathrm{SiO_2}\) cladding, and \({C_w}\) represents the width of the entire cladding. The entire \(\mathrm{Si_3N_4}\) waveguide is encapsulated within \(\mathrm{SiO_2}\). \({C_w}\) should be a large value, but considering the needs of finite element simulation, a relatively large value is chosen for \({C_w}\) in the actual simulation. Under the mode supported by the waveguide, the energy will hardly leak into the \(\mathrm{SiO_2}\).


 The design of the straight waveguide is the same as that of the microring to ensure that both the microring cavity and the straight waveguide support the same spatial mode.
The cross-sectional side view of the resonator's architecture is depicted in Figure~\ref{fig:C0}~(a)~\ding{192}, where the gap $d$ is defined as the shortest straight-line distance at half the height ($h/2$) between the two waveguides. As light propagates along the waveguides in Figure~\ref{fig:C0}~(a), effective mode area $A_{{eff}}$ represents the effective cross-sectional area of the waveguide,\cite{dudley2006supercontinuum} 
\begin{equation}\label{Aeff}
A_{{eff}}=\frac{\left(\iint_{-\infty}^{+\infty}|F(x, y)|^{2} d x d y\right)^{2}}{\iint_{-\infty}^{+\infty}|F(x, y)|^{4} d x d y},
\end{equation}
where $F(x, y)$ is the mode distribution in $\mathrm{Si_3N_4}$ and $\mathrm{SiO_2}$, assuming the mode distribution within the resonator remains constant over time.

\begin{figure}
  \includegraphics[width=0.6\linewidth]{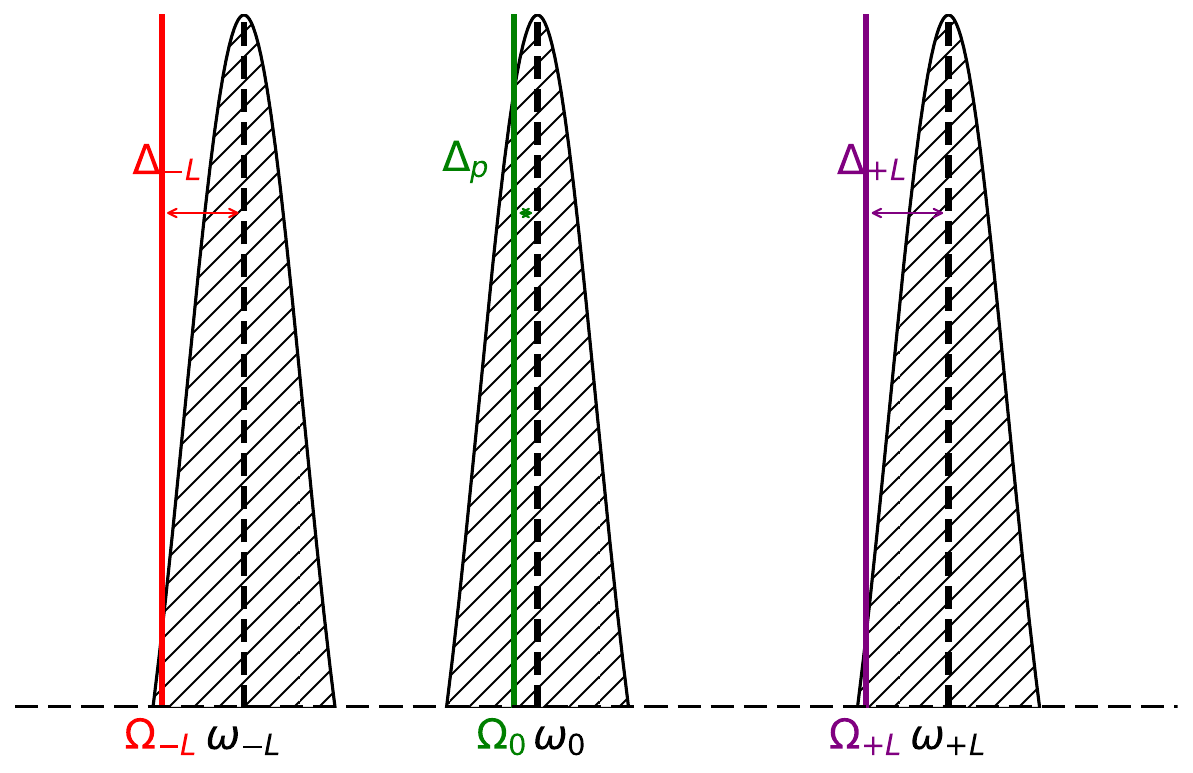}
  \caption{Resonances under anomalous dispersion. Solid lines are perfectly equispaced, which represents the location of the output comb lines. Black slashes represent the location of the resonances under anomalous dispersion. It illustrates the relationship between the laser frequency $\Omega_p=\Omega_0$, cold-resonance frequency $\omega_p=\omega_0$, and the cold cavity pump detuning $\Delta_p =\omega_0-\Omega_0$. As for mode $-L/+L$, the cold cavity detuning $\Delta_{-L/+L}=\omega_{-L/+L}-\Omega_{-L/+L}=\Delta_p + D_{int}$, assuming $D_2\gg D_3$ and $L$ small.}
  \label{fig:S0}
\end{figure}

The geometry of the coupling region is crucial in determining the ring and straight waveguide coupling rate, facilitating the extraction of the coupling rate, which describes the resonator's input-output relationship.\cite{breunig2016three} We focus on the fundamental $\mathrm{TE_{00}}$ mode as an example, resulting in a Lorentzian-shaped envelope for cavity resonance (illustrated in Figure~\ref{fig:S0}, shaded area). Due to the refractive index $n(\omega)$ of the material being frequency-dependent, resonances do not appear evenly spaced in the spectrum. Adjusting cavity variables such as detuning, dispersion, coupling, and loss rate can tune our system to different work points.

\subsection{Detuning, Dispersion, and Cavity Structure}
This section explores the properties of detuning and dispersion in relation to the cavity structure. The relative mode number \( L \) (\( L \in \mathbb{Z} \)) is introduced to define the state modes alongside the pump mode \( \omega_0 \) (\( L=0 \)). The resonance modes around \( \omega_0 \) can be described using a Taylor expansion:
\begin{equation}\label{Taylor}
	\omega_L = \omega_0 + D_1 L + \frac{D_2}{2} L^2 + \cdots = \omega_0 + \sum_{{n}=1}^{\infty} D_{{n}} \frac{L^{n}}{{n}!},
\end{equation}
with \(D_n = \left. \frac{d^n \omega_L}{d L^n} \right|_{L=0}\). Notably, the term \(\frac{D_1}{2\pi}\) denotes the comb's free spectral range (FSR, $fsr$). The coefficient \(D_2\) pertains to group velocity dispersion (GVD), with \(D_2 > 0\) signaling anomalous dispersion and \(D_2 < 0\) indicating normal dispersion. For higher orders (\(n > 3\)), we simplify the model by setting \(D_n = 0\), thus the integrated dispersion \(D_{{int}}= \omega_L-\omega-D_1 L = \frac{D_2}{2} L^2 + \frac{D_3}{6} L^3\). Under appropriate microcavity structure design, $D_2 \gg D_3$, $D_{int}$ closely approximates a quadratic polynomial near \(\omega_0\) when $L$ is not very large, as shown in Figure~\ref{fig:C0}~(b, c).

At the below-threshold region, where pump power is weak, we can neglect frequency shifts caused by the pump power through nonlinear effects, called the ``cold cavity''. The pump detuning is given by:
\begin{equation}
	\Delta_p = \omega_0 - \Omega_0.
\end{equation}
where $\omega_0$ is the resonance peak of the pump mode and $\Omega_0$ is the pump light frequency. The generated QFC teeth mode detuning follows:
\begin{equation}
	\Delta_L = \omega_L - \Omega_L.
\end{equation}
For QFCs, the quantum noise is centered around $\omega_0 \pm n D_1$ , just like the classical field envelope, $\Delta_L = \Delta_p + D_{int}= \Delta_p+\frac{D_2}{2} L^2+ \frac{D_3}{6} L^3$.

\subsection{Coupling, Loss, and $\mathbf{gap}$}
This section delves into the interplay between the input-output parameters, specifically the coupling rate ($\gamma$) and loss rate ($\mu$) of the cavity, along with the intrinsic cavity parameters: gap ($d$) and quality factors ($Q$). The ratio $r = \gamma / \mu$ serves as a measure of the relationship between coupling and intrinsic loss. A value of $r < 1$ indicates under coupling, $r > 1$ signifies over coupling, and $r = 1$ denotes critical coupling. The overall damping rate ($\Gamma$) is defined as the sum of the coupling and loss rates:
\begin{equation}
    \Gamma = \gamma + \mu.
\end{equation}
$\Gamma$ approximates the full width at half maximum (FWHM) of the resonance, following the relation $\Gamma = \frac{\omega}{Q}$. The quality factor is composed of the intrinsic quality factor ($Q_0 = \frac{\omega}{\mu}$) and the external quality factor ($Q_{ex} = \frac{\omega}{\gamma}$). The total quality factor $Q$ is given by:
\begin{equation}
    \frac{1}{Q} = \frac{1}{Q_0} + \frac{1}{Q_{ex}}.
\end{equation}

\subsection{Actual Simulation}
Our simulation parameters are presented in Table~\ref{tab:tableC1}. The radius of the microring cavity is designed to be \(240 \, \mathrm{\mu m}\), with a FSR of approximately \(100 \, \mathrm{GHz}\). In the frequency range of \(130 \, \mathrm{THz}\) to \(260 \, \mathrm{THz}\), the microcavity supports a total of four spatial modes: \(\mathrm{TE_{00}}\), \(\mathrm{TM_{00}}\), \(\mathrm{TE_{10}}\), and \(\mathrm{TM_{10}}\). Their mode profiles are shown in the right-hand figure of Figure~\ref{fig:C0}~(a)~\ding{193}, where it can be seen that the energy of the optical field is well confined within the waveguide structure. By simulating these four modes, with the nearest resonance peak near \(214.6 \, \mathrm{THz}\) as the zero-dispersion point for each modal family $\omega_0$, we can plot the integrated dispersion ($D_{{int}}$) curve as shown in Figure~\ref{fig:C0}~(b, c). For each modal family, the detailed parameters are given in Table~\ref{tab:tableC2}. As shown in Figure~\ref{fig:C0}~(b), the integrated dispersion curve is not a constant line, so the resonance peaks in the microcavity are not equally spaced. Our four modal families are all anomalous dispersion around \(214.6 \, \mathrm{THz}\). ${D_1}$, ${D_2}$, ${D_3}$, ${D_4}$, and ${D_5}$ are the first, second, third, fourth, and fifth order dispersion coefficients. In Table~\ref{tab:tableC2}, we can see ${D_1}$ and ${D_2}$ are much larger than the others, so the $D_{{int}}$ curve in Figure~\ref{fig:C0}~(c) appears as a parabola when $l$ is not very large. The FSR equals to ${\frac{D_1}{2 \pi}}$, ${f_0}$ and ${\lambda_0}$ are the frequency and wavelength of the zero-dispersion point. \(n_{{eff}}\) is the effective refractive index, and \(g_0\) is the nonlinear coupling rate. The total quality factor $Q$, according to Wu's thesis,\cite{alma991012980422503412} is set at $10^6$ for $\mathrm{TE_{00}}$ and $\mathrm{TM_{00}}$, and $5 \times 10^5$ for $\mathrm{TE_{10}}$ and $\mathrm{TM_{10}}$.

\begin{figure}
  \includegraphics[width=0.6\linewidth]{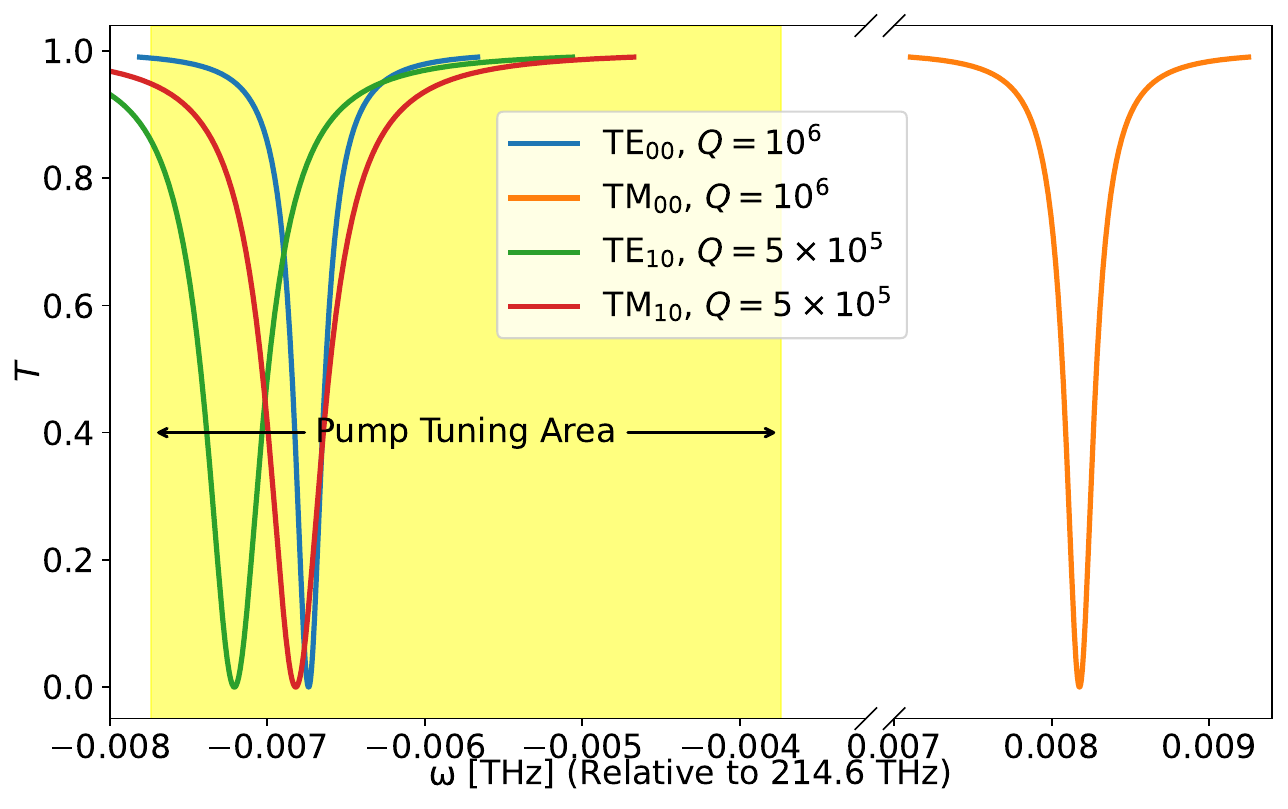}
  \caption{\label{fig:C1}Transmission spectrum of our cavity around $\mathrm{214.6~THz}$. $T$ is the aligned transmission coefficient. Different modal families experiencing different Q factors are plotted in Lorentzian line shapes ranging ${5\Delta_\omega}$ with different colors (blue: $\mathrm{TE_{00}}$, orange: $\mathrm{TM_{00}}$, green: $\mathrm{TE_{10}}$, red: $\mathrm{TM_{10}}$), ${\Delta_\omega}$ is the full width at half maximum of the resonators. $\mathrm{TE_{00}}, \mathrm{TE_{10}}, \mathrm{TM_{10}}$ have a significant overlap region, allowing for effective excitation of all three resonance modes using the same frequency pump light. The yellow region in the background represents the tuning range of the pump light frequency, spanning approximately 4 GHz across the entire selection area.}
\end{figure}

To simplify the pumping conditions in our design for achieving spatial multiplexing quantum entanglement generation using WGMRs, we employ monochromatic pumping light to simultaneously pump all spatial modes. To ensure that the pumping light is nearly resonant with the peaks of the all spatial modes, we traverse all the resonance peaks, aiming to find as many overlapping resonance peaks as possible. Near \(214.6 \, \mathrm{THz}\), we devised Figure~\ref{fig:C1}, showing that the resonances of $\mathrm{TE_{00}}$, $\mathrm{TE_{10}}$ and $\mathrm{TM_{10}}$ are mostly overlapped. We can tune our pump light around the light yellow area to examine the QFCs generation. The key aspect of this design lies in ensuring the close matching of the pumping light's frequency with those of the three spatial modes, thereby achieving optimal excitation and laying a solid foundation for simultaneous quantum entanglement generation via optical frequency comb preparation.
We employed a sophisticated technique involving the utilization of an SLM. Through precise manipulation facilitated by the SLM, we could dynamically adjust the spatial modes of the pumping light. This dynamic adjustment enabled us to tailor the pump beam to resonate effectively with all three spatial modes simultaneously.

Considering the intricate nature of our system, it was imperative to account for the unique characteristics of each spatial mode. In our setup, we harnessed three distinct spatial modes: $\mathrm{TE_{00}}$, $\mathrm{TE_{10}}$, and $\mathrm{TM_{10}}$. Each of these modes possesses its own resonant frequency and spatial distribution within the WGMR.

\begin{figure}
  \includegraphics[width=0.6\linewidth]{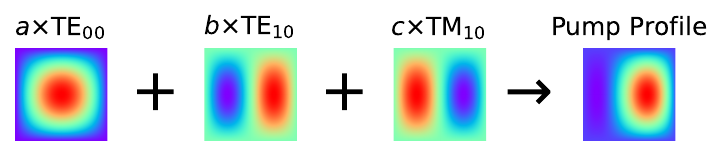}
  \caption{\label{fig:E1}Pump mode profile modulation. To enable using pump light with same frequency to excite all three modal families, the mode profile of the pump light needs to be a superposition of the three modal families.}
\end{figure}

In essence, the pumping light needed to be a composite of these three spatial modes, with each mode contributing to the overall excitation process. This composite nature of the pumping light was aptly depicted in Figure~\ref{fig:E1}, illustrating the superposition of $\mathrm{TE_{00}}$, $\mathrm{TE_{10}}$, and $\mathrm{TM_{10}}$ modes. Crucially, the relative amplitudes of these modes, depicted by proportionality coefficients in the figure, determined the contribution of each mode to the overall pumping process.

By meticulously crafting the pumping light to encompass the superposition of these spatial modes, we ensured comprehensive excitation coverage within the WGMR. This approach does not need to tune pump frequency using 
electro-optic modulator or other pump source, only use SLM, realizing simultaneously pumping all spatial modes.

\section{$\mathbf{\chi^{(3)}}$ WGMRs Model Supporting Multiple Modal Families}
\subsection{Hamiltonian and Dynamics}
A general WGMR without nonlinearity supporting multiple modal families can be modeled as a series of resonance peaks of frequencies in distinct modal families. Such linear WGMRs has the following Hamiltonian:
\begin{equation}\hat{H}_0=\sum_{i,j} \hbar \omega_{i,j}\hat{a}^\dagger_{i,j} \hat{a}_{i,j},\end{equation}
where $\omega_{i,j}$ represents the resonances peak frequency of the $j$-th resonance in the $i$-th modal family, $\hbar$ is the reduced Planck constant, and $\hat{a}_{i,j}$ and $\hat{a}_{i,j}^\dagger$ are the annihilation and creation operators of $j$-th resonance mode in the $i$-th modal family. 

This Hamiltonian is usually called the ``free Hamiltonian'' because it does not describe any interaction between modes. In $\mathrm{Si_3N_4}$ WGMRs, mode interaction is introduced by the Kerr effect, enabling SPM, XPM, spontaneous and stimulated FWM and Bragg scattering. These nonlinearities can be modeled as an additional term in the free Hamiltonian: 
\begin{equation}\hat{H}_{int}=-\frac{1}{2}\sum_i \hbar \eta_i \sum_{j_1+j_2=j_3+j_4} \hat{a}_{i,j_1}^\dagger\hat{a}_{i,j_2}^\dagger \hat{a}_{i,j_3}\hat{a}_{i,j_4}. \end{equation}
The Kerr nonlinearity is a third-order nonlinear effect that describes the interaction between four electromagnetic modes. The nonlinear process must also satisfy the principle of energy conservation, so the cavity modes involved in the Kerr nonlinearity are subject to the constraint \(j_1 + j_2 = j_3 + j_4\).
\(\eta_i\) corresponds to the nonlinear coupling coefficient of the \(i\)-th modal family. In this paper, we assume that all third-order nonlinear processes generated within the same modal family have the same nonlinear coupling coefficient. Here, the lower bound estimate of the nonlinear coupling coefficient is \( \eta_i = \frac{\hbar \omega_0^2 c n_2}{n_0^2 V_{{eff}}} \) \cite{chembo2016quantum}, representing the per photon frequency shift of the resonance due to the ${\chi}^{(3)}$ nonlinearity. Here, \(c\) is the vacuum speed of light, and \(n_2\) is the nonlinear index of $\mathrm{Si_3N_4}$, which is associated with the refractive index \(n_0\). In our $\mathrm{Si_3N_4}$ microresonator, \(n_2 = 2.6 \times 10^{-19} \, \mathrm{m^2~W^{-1}}\).\cite{wang2021integrated} The effective mode volume \( V_{\mathrm{eff}} \) can be defined as:\cite{gao2022probing}

\begin{equation}
    V_{{eff}} = \frac{\int n_0^2 |F(x,y,z)|^2 \, dV \int |F(x,y,z)|^2 \, dV}{\int n_0^2 |F(x,y,z)|^4 \, dV}.
\end{equation}

When the WGMR resonator is a microring resonator, the upper bound estimate of \( V_{{eff}} \) can be approximated by \( V_{{eff}} \approx A_{{eff}} \cdot 2\pi R \).\cite{chembo2016quantum}

The total Hamiltonian is then given by:
\begin{equation}
    \hat{H}=\hat{H}_{0}+\hat{H}_{int},
\end{equation}
which is the combination of the free and interacting Hamiltonian. The process described by this Hamiltonian is relatively comprehensive, and considering the interactions of multiple modes makes it very challenging to study. To address this issue, we consider specific situations and reasonably neglect certain processes to simplify the calculations.

In this paper, we primarily focus on microcavity systems with below-threshold pumping, assuming there is no cross modal families interaction. The new Hamiltonian can be written as a sum of the Hamiltonian of distinct modal families: 
\begin{equation}
\hat{H}_{int}=\sum_{i}\hat{H}_{i}.\end{equation}
\begin{equation}
\label{eq:hamil}
\begin{aligned}
   \hat{H}_{i}=&-\frac{1}{2}\hbar\eta_i \hat{a}_{i,p}^\dagger\hat{a}_{i,p}^\dagger\hat{a}_{i,p} \hat{a}_{i,p}- \sum_L^N \hbar \eta_{i} [ \frac{1}{2}(\hat{a}_{i,-L}^\dagger\hat{a}_{i,-L}^\dagger\hat{a}_{i,-L} \hat{a}_{i,-L}+\hat{a}_{i,+L}^\dagger\hat{a}_{i,+L}^\dagger\hat{a}_{i,+L} \hat{a}_{i,+L} )\\&+2(\hat{a}_{i,p}^\dagger\hat{a}_{i,-L}^\dagger\hat{a}_{i,p} \hat{a}_{i,-L}+\hat{a}_{i,p}^\dagger\hat{a}_{i,+L}^\dagger\hat{a}_{i,p} \hat{a}_{i,+L}+\hat{a}_{i,-L}^\dagger\hat{a}_{i,+L}^\dagger\hat{a}_{i,-L} \hat{a}_{i,+L}) +(\hat{a}_{i,-L}^\dagger\hat{a}_{i,+L}^\dagger\hat{a}_{i,p} \hat{a}_{i,p} +\hat{a}_{i,p}^\dagger\hat{a}_{i,p}^\dagger\hat{a}_{i,-L} \hat{a}_{i,+L} ) ] .
\end{aligned}
\end{equation}
Here, the total mode number in each modal family is set to $2N+1$. Apart from the pump light mode, we also consider \(N\) pairs of frequency modes near the pump light frequency. The first term corresponds to the self-phase modulation of the \(i\)-th modal family pump light. The second term is the self-phase modulation effect of the frequency components \(+L\) and \(-L\). The third term is the cross-phase modulation effect between the frequency components \(+L\), \(-L\) and the pump light. The last term describes the process of the pump light generating the frequency components \(+L\) and \(-L\). In the above formulation of the Hamiltonian, we have not considered the Bragg scattering effects and other four-wave mixing processes that occur at above-threshold pumping condition. These approximations allow the \(+L\) and \(-L\) frequency components to be considered as evolving together and not affected by other mode pairs, which is reasonable for situations slightly above the threshold. Although we consider the quantum entangled optical frequency combs to be below the threshold, it is also important to introduce above-threshold effects (SPM and XPM) to determine the threshold value.

This comprehensive Hamiltonian framework underscores the intricate quantum dynamics at play within the resonator, laying the groundwork for understanding and optimizing the generation of entangled photon pairs through the pump-degenerate FWM process.

We can then employ the Heisenberg-Langevin formalism to model the dynamics of all interacting modes within a resonator. A typical linear Heisenberg-Langevin equation has the following form:\cite{debuisschert1993type}
\begin{equation}
    \frac{\mathrm{d} \hat{a}}{\mathrm{d} t} =  -\Gamma \hat{a}-\mathrm{i} \Delta \hat{a}+\sqrt{2\gamma} \hat{a}^{\mathrm{in}}+\sqrt{2\mu} \hat{a}^{\mathrm{loss}},
\end{equation}
where the first term on the right is damping, describing the leakage of the electric field in the WGMR to the environment, including the coupling waveguide and the bath. The second term describes the oscillation of the electric field within the cavity, with $\Delta = \omega -\Omega$ being the detuning of the pumping frequency $\Omega$ from the pumping resonance frequency $\omega$. The last two terms describe the transfer of the waveguide optical field and the bath optical field into the cavity. The damping rate(total loss rate) $\Gamma$ equals to the sum of the external coupling rate $\gamma$ and the intrinsic loss rate $\mu$.

As previously explained, the \(+L\) and \(-L\) modes in each modal family can be considered to change only with the pump light under below-threshold and slightly above-threshold pumping conditions. The evolution of the pump, \(+L\), and \(-L\) modes is governed by the Heisenberg-Langevin equations with nonlinear coupling added to the linear one. The corresponding Heisenberg-Langevin equations for these modes are as follows:
\begin{equation}
    \label{Heisenberg-Langevin equation}
\left\{
\begin{aligned}
    \frac{\mathrm{d} \hat{a}_{i,p}}{\mathrm{d} t} ~=~&  \mathrm{i} \eta_{i, p}\left[\left(\hat{a}_{i,p}^{\dagger} \hat{a}_{i,p}+2 \hat{a}_{i,-L}^{\dagger} \hat{a}_{i,-L}+2 \hat{a}_{i,+L}^{\dagger} \hat{a}_{i,+L}\right) \hat{a}_{i,p}+2 \hat{a}_{i,p}^{\dagger} \hat{a}_{i,-L} \hat{a}_{i,+L}\right] -\Gamma_i \hat{a}_{i,p}-\mathrm{i} \Delta_{p} \hat{a}_{i,p}\\&+\sqrt{2\gamma_i} \hat{a}_{i,p}^{\mathrm{in}}+\sqrt{2\mu_i} \hat{a}_{i,p}^{\mathrm{loss}}, \\
    \frac{\mathrm{d} \hat{a}_{i,-L}}{\mathrm{d} t} ~=~& \mathrm{i} \eta_{i, -L} \left[\left(2 \hat{a}_{i,p}^{\dagger} \hat{a}_{i,p}+\hat{a}_{i,-L}^{\dagger} \hat{a}_{i,-L}+2 \hat{a}_{i,+L}^{\dagger} \hat{a}_{i,+L}\right) \hat{a}_{i,-L}+\hat{a}_{i,p}^{2} \hat{a}_{i,+L}^{\dagger}\right]-\Gamma_i \hat{a}_{i,-L}-\mathrm{i} \Delta_{-L} \hat{a}_{i,-L}\\
    &+\sqrt{2\gamma_i} \hat{a}_{i,-L}^{\mathrm{in}}+\sqrt{2\mu_i} \hat{a}_{i,-L}^{\mathrm{loss}}, \\
    \frac{\mathrm{d} \hat{a}_{i,+L}}{\mathrm{d} t} ~=~& \mathrm{i} \eta_{i, +L}\left[\left(2 \hat{a}_{i,p}^{\dagger} \hat{a}_{i,p}+2 \hat{a}_{i,-L}^{\dagger} \hat{a}_{i,-L}+\hat{a}_{i,+L}^{\dagger} \hat{a}_{i,+L}\right) \hat{a}_{i,+L}+\hat{a}_{i,p}^{2} \hat{a}_{i,-L}^{\dagger}\right] -\Gamma_i \hat{a}_{i,+L}-\mathrm{i} \Delta_{+L} \hat{a}_{i,+L}\\
    &+\sqrt{2\gamma_i} \hat{a}_{i,+L}^{\mathrm{in}}+\sqrt{2\mu_i} \hat{a}_{i,+L}^{\mathrm{loss}}.
\end{aligned}
\right.
\end{equation}
Here, $\Gamma_i$ represents the total loss rate in $i$-th modal family, which is a combination of the external coupling rate $\gamma_i$ and the intrinsic loss rate $\mu_i$. The operators $\hat{a}_{i,\pm L}^{\mathrm{in}}$ and $\hat{a}_{i,\pm L}^{\mathrm{loss}}$ represent the incoming driving fields and the fields lost from the resonator, respectively, with statistical properties reflecting vacuum fluctuations and coherent pump inputs. In the following discussion, we will omit the spatial mode index \(i\) for convenience.

\subsection{Steady-state Equations}
The temporal evolution of cavity modes is governed by Equation~(\ref{Heisenberg-Langevin equation}). To find the solutions, we employ the linearization method by expanding each field operator $\hat{a}_{j}$ into its steady-state mean value $\alpha_{j}$ and fluctuation operator $\delta \hat{a}_{j}$, such that $\hat{a}_{j} = \alpha_{j} + \delta \hat{a}_{j}$. In the steady state, $\alpha_{j}$ remains constant, so by setting $\delta \hat{a}_{j} = 0$ and $\mathrm{d} \alpha_{j} / \mathrm{d} t = 0$, we can derive the Heisenberg-Langevin equations in the steady state. Under these conditions, the inputs of the signal and idler, as well as the losses of the signal, idler, and pump, are all in the vacuum state, implying $\alpha_{s}^{\mathrm{in}} = \alpha_{i}^{\mathrm{in}} = \alpha_{j}^{\mathrm{loss}} = 0$.

For simplicity, we take the phase of the external pump as a reference. Therefore, we set $\alpha_{j} = A_{j} \mathrm{e}^{\mathrm{i} \theta_{j}}$, $\alpha_{p}^{\mathrm{in}} = A^{\mathrm{in}} \mathrm{e}^{\mathrm{i} \theta_{\mathrm{in}}}$, $\phi = \theta_{-L} + \theta_{+L} - 2 \theta_{p}$, and $\psi = \theta_{\mathrm{in}} - \theta_{p}$. We assume $A_{-L} = A_{+L} = A$ and $\Delta_{+L} = \Delta_{-L} = \Delta$. The external pump power is defined as
$F = \sqrt{\frac{2 \gamma \eta}{\hbar \Omega_0 \Gamma^3} P_{\mathrm{in}}}$, with $A^{\mathrm{in}} = F \sqrt{\frac{\Gamma^3}{2 \gamma \eta}}$.\cite{matsko2005optical,gonzalez2017third} Using these variables, we derive the following equations:

\begin{equation}\label{steady-state solution}
	\begin{cases}
	A_p^4 = 1 + \left(\Delta_p - \frac{D_3}{2} - 2A_p^2 - 3A^2\right)^2 \\
	F^2 = A_p^2\left\{\left(1 + 2\frac{A_p^2}{A_p^2}\right)^2 + \left[\Delta_p - A_p^2 - \frac{2A^2}{A_p^2}\left(\Delta_p - \frac{D_3}{2} - 3A^2\right)\right]^2\right\} \\
	\sin(\phi) = \frac{1}{A_p^2} \\
	\cos(\phi) = \frac{\Delta - \left(2A_p^2 + 3A_p^2\right)}{A_p^2} = \frac{1}{A_p^2}\left(\Delta_p - \frac{D_3}{2} - 3A^2 - 2A_p^2\right) \\
	\sin(\psi) = \frac{A_p}{F}\left[\Delta_p - A_p^2 - \frac{2A^2}{A_p^2}\left(-3A^2 - \frac{D_3}{2} + \Delta_p\right)\right] \\
	\cos(\psi) = \frac{A_p}{F}\left(1 + \frac{2A^2}{A_p^2}\right)
	\end{cases}.
\end{equation}

By assigning specific values to any two of ${A}_{p}$, ${A}$, ${A}^{\mathrm{in}}$, ${\Delta_p}$, and ${\Delta}$, we can numerically determine the relationship between the remaining three variables, enabling us to express any two of them as a function of the third.

\subsection{Quantum Fluctuation Equations}
To explore the quantum properties of the signal and idler, we need the quantum fluctuation equations derived from Equation~(\ref{Heisenberg-Langevin equation}). Given that we have the steady-state equations, we modify Equation~(\ref{Heisenberg-Langevin equation}) by setting $\hat{a}_{j} = \alpha_j + \delta \hat{a}_{j}$. We treat the pump field as a classical field, thus $\delta\hat{a}_{p} = 0$. Higher-order fluctuations are also neglected.

Define a vector containing the fluctuations of the signal and idler as:
\begin{equation}
	\delta \hat{\boldsymbol{A}} = \left(\delta \hat{a}_{-L} \mathrm{e}^{-\mathrm{i} \theta_{-L}}, \delta \hat{a}_{-L}^{\dagger} \mathrm{e}^{\mathrm{i} \theta_{-L}}, \delta \hat{a}_{+L} \mathrm{e}^{-\mathrm{i} \theta_{+L}}, \delta \hat{a}_{+L}^{\dagger} \mathrm{e}^{\mathrm{i} \theta_{+L}}\right)^{\mathrm{T}},
\end{equation}
where $\theta_{j}$ is the phase of the mean value $\alpha_{j} = A_{j} \mathrm{e}^{\mathrm{i} \theta_{j}}$. The time evolution of these fluctuations is given by:
\begin{equation}
	\frac{1}{\Gamma} \cdot \frac{\mathrm{d} \delta \hat{\boldsymbol{A}}}{\mathrm{d} t} = M \cdot \delta \hat{\boldsymbol{A}} + T^{\mathrm{in}} \cdot \delta \hat{\boldsymbol{A}}^{\mathrm{in}} + T^{\mathrm{loss}} \cdot \delta \hat{\boldsymbol{A}}^{\mathrm{loss}},
\end{equation}
where $\Gamma=\Gamma_i$, $T^{\mathrm{in}} = \mathrm{diag}\left(\sqrt{2{\gamma}}, \sqrt{2{\gamma}}, \sqrt{2{\gamma}}, \sqrt{2{\gamma}}\right)$ and $T^{\mathrm{loss}} = \mathrm{diag}\left(\sqrt{2{\mu}}, \sqrt{2{\mu}}, \sqrt{2{\mu}}, \sqrt{2{\mu}}\right)$. The matrix $M$ is derived from the linearization process, and its elements are related to the field's mean values and detunings.

The evolution of these fluctuations in the frequency domain can be given by the Fourier transform. After introducing the input-output relationship of the resonator:
\begin{equation}
	\hat{a}^{\mathrm{out}} = -\hat{a}^{\mathrm{in}} + \sqrt{2\gamma} \hat{a},
\end{equation}
we obtain:
\begin{equation}
	\begin{aligned}
		\delta \hat{\boldsymbol{A}}^{\mathrm{out}}(\omega) & = -\delta \hat{\boldsymbol{A}}^{\mathrm{in}} + T \delta \hat{\boldsymbol{A}} \\
		& = \left(T\left(\mathrm{i}\omega-M\right)^{-1}T^\mathrm{in}-I\right)\cdot\delta\hat{\boldsymbol{A}}^{in}+T\left(\mathrm{i}\omega-M\right)^{-1}T^\mathrm{loss}\cdot\delta\hat{\boldsymbol{A}}^\mathrm{loss}
	\end{aligned},
\end{equation}
where $T = \mathrm{diag}\left(\sqrt{2\gamma}, \sqrt{2\gamma}, \sqrt{2\gamma}, \sqrt{2\gamma}\right)$,
and $I$ is the identity matrix.

The output spectral noise density matrix is defined by:
\begin{equation}
	\begin{aligned}
		S(\omega) =& \left\langle \delta \hat{\boldsymbol{A}}^{\mathrm{out}}(\omega) \delta \hat{\boldsymbol{A}}^{\mathrm{out}~ \mathrm{T}}(-\omega) \right\rangle \\
		 =& \left(T\left(\mathrm{i}\omega-M\right)^{-1}T^\mathrm{in}-I\right)\langle\delta\hat{\boldsymbol{A}}^\mathrm{in}(\omega)\delta\hat{\boldsymbol{A}}^{\mathrm{in~T}}(-\omega)\rangle\left(T\left(-\mathrm{i}\omega-M\right)^{-1}T^\mathrm{in}-I\right)^T\\&+\left(T\left(\mathrm{i}\omega-M\right)^{-1}T^\mathrm{loss}\right)\langle\delta\hat{\boldsymbol{A}}^\mathrm{loss}(\omega)\delta\hat{\boldsymbol{A}}^{\mathrm{loss~T}}(-\omega)\rangle\left(T\left(-\mathrm{i}\omega-M\right)^{-1}T^\mathrm{loss}\right)^T
	\end{aligned}.
\end{equation}

\section{Bipartite Entanglement Generation and Analysis in WGMRs}
When a monochromatic pump light is below the threshold, it generates bipartite entanglement on both sides symmetric to the pump frequency. As the pump light power increases, it will excite other resonances above the threshold, which will act as new pump lights, generating a complex entanglement structure. For bipartite entanglement in Gaussian continuous-variable quantum physics, we can use the Duan criterion to distinguish the degree of quantum entanglement.\cite{ferrini2014symplectic,giovannetti2003characterizing,gonzalez2017third,duan2000inseparability} 

Here, we define the position operator and momentum operator as $\hat{X}_{1,2}=\frac{\hat{a}_{1,2}+\hat{a}^\dagger_{1,2}}{\sqrt{2}}$, $\hat{Y}_{1,2}=\frac{-\mathrm{i} (\hat{a}_{1,2}-\hat{a}^\dagger_{1,2})}{\sqrt{2}}$. The rotated position and momentum operators are the superposition of the position and momentum operators, \begin{equation}
\left(\begin{array}{c}Y_\pm^{rot}\\X_\pm^{rot}\end{array}\right)=\left(\begin{array}{cc}\cos(\theta_\pm)&\sin(\theta_\pm)\\-\sin(\theta_\pm)&\cos(\theta_\pm)\end{array}\right)\left(\begin{array}{c}  
\frac{\hat{Y}_1\pm\hat{Y}_2}{\sqrt{2}} \\\frac{\hat{X}_1\pm\hat{X}_2}{\sqrt{2}}\end{array}\right).
\end{equation}
The Duan criterion has the following form:
\begin{equation}
    {C}=\Delta^2 (\hat{X}_-^{rot}) + \Delta^2 (\hat{Y}_+^{rot}) - |G| \geq 0,
\end{equation}
where $G=cos(\theta_+-\theta_-)$.

In above equation, $\Delta^2 (*)$ represents the covariance of operators, and it is the superposition of elements of $S(\omega)$. 
 If the Duan criterion is not satisfied, that is ${C}<0$, the bipartite elements are entangled, and the smaller the value of ${C}$, the better the quantum entanglement. We can tune $\theta_+$ and $\theta_-$ to minimize ${C}$ to ${C_{min}}$, which is the real entanglement degree hidden behind specific operators.

\section{Results}

\textcolor{black}{
\subsection{Below-Threshold Stability}
In this section, we will delve into the solutions of the steady-state equation corresponding to Equation 16 below the threshold and analyze the stability of these solutions. To obtain the below-threshold solutions, we set the amplitude terms \(A\) for the signal and idler light in the steady-state equation to zero. This method allows us to isolate the solutions that exist below the threshold. In Figure~\ref{fig:Q0}~(a), we plotted the quantum  entanglement degrees ($C_{min}$) corresponding to these solutions by selecting pump light amplitude (\(A_p\)) values between \(0\) and \(1 \times 10^{10} \, \mathrm{V/m}\), and pump light frequency detuning (\(\Delta_p\)) values ranging from \(-1 \) to \(3 \, \mathrm{GHz}\). However, it's important to note that these solutions are not necessarily stable.\\
The instability of the below-threshold solutions corresponds to the regions above the threshold. The stability of the solutions can be determined by examining the following stability condition equation: 
\begin{equation}
M < 0.
\end{equation}
Here, \(M\) is the matrix that describes the evolution of the fluctuations in the signal and idler light modes below the threshold (Equation~25),
\begin{equation}
    	\frac{\mathrm{d} \delta \hat{\boldsymbol{A}}}{\mathrm{d} t} = M \cdot \delta \hat{\boldsymbol{A}}.
\end{equation}
The condition \(M < 0\) means that all the eigenvalues of the \(M\) matrix must be negative. If any of the eigenvalues become positive, the below-threshold solution loses its stability and will eventually exceed the threshold, leading to a different regime of operation.\\
By solving this stability equation, we can delineate the regions where the below-threshold solutions become unstable. These regions are bounded by the threshold, and we have marked them in Figures~\ref{fig:Q0},~\ref{fig:Q1}, and~\ref{fig:Q2}. \\
References\cite{ng_quantum_2023,vendromin_highly_2024} provide critical insights into the behavior of the system as it approaches and crosses the threshold.
According to them, near the threshold, linearization method breaks down, meaning that Equation~18 no longer holds in this region. This failure occurs because linearized approximations are unable to fully capture the complex nonlinear dynamics near the threshold, where classical and quantum effects becomes closer. Consequently, for regions closer to the threshold, where the linearized theory may fail, the system behavior becomes more challenging to predict accurately using simple analytical models.\\
On the other hand, for regions far from the threshold and well below it, the linearized model works quite well. This is in line with the findings of studies such as the reference,\cite{guidry2022quantum} which combined both experimental and theoretical investigations of the quantum properties of comb lines during the generation of soliton frequency combs. Using single-photon counting techniques, the study demonstrated that the linearized theoretical model aligned remarkably well with experimental data for dissipative Kerr soliton modes below the threshold. This alignment lends credibility to the linearized approach for studying below-threshold dynamics, provided the system is sufficiently far from the threshold.\\
Therefore, in our optimization process, we consciously avoided regions near the threshold that could potentially lead to the breakdown of the linearized model. This deliberate choice ensures that our model remains accurate and reliable, and it simplifies the analysis of quantum optical combs in the below-threshold regime. By carefully navigating around the regions where the linearized theory is likely to fail, we can confidently explore the parameter space and ensure the stability of the solutions we study.
} 

\subsection{Phase Diagram of Bipartite Entanglement in Single Modal Family}
\begin{figure}
  \includegraphics[width=\linewidth]{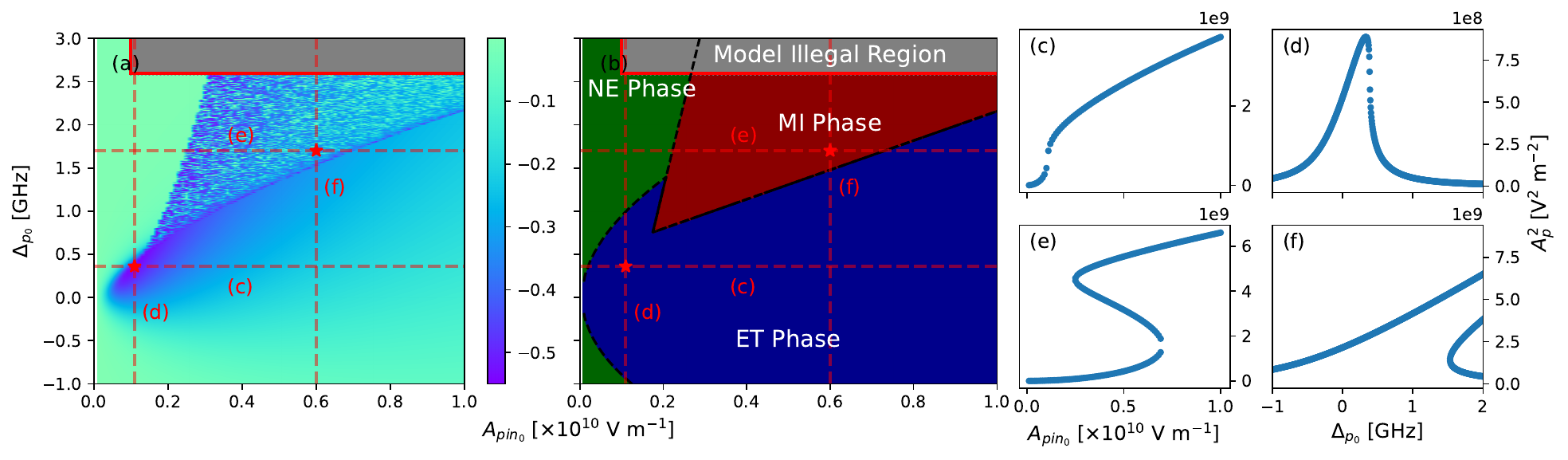}
  \caption{\label{fig:Q0}(a) Heat map of the relationship between the quantum entanglement degree (${C_{min}}$), pump light frequency (${\Delta_{p_0}}=\omega_{p_0}-\Omega_{p_0}$), and pump light amplitude (${A_{pin_0}}$) in the fundamental mode $\mathrm{TE_{00}}$, where color indicates the value of ${C_{min}}$. The two star-marked points here are at the same position of (b), \textcolor{black}{gray region with red edges: invalid regions where linearization breaks down, the same in (b)}. (b) Phase diagram corresponding to (a). Based on the characteristics of the entanglement degree, the entire diagram is divided into three phases: the Non-Entanglement (NE) Phase with entanglement degree almost zero, the Entanglement-Tunable (ET) Phase with entanglement degree gradually changing, and the Modulation Instability (MI) Phase with entanglement degree unstable. One of the chosen two points is in ET Phase, the other MI Phase. (c, d) For the point in the ET Phase, we conducted cross-sectional scans along their corresponding pump frequencies or amplitudes to study the characteristics of different phases and the variation in the power of the intra-cavity pump light during the phase transition process. (e, f) The same process as (c, d), but point in MI Phase.}
\end{figure}

In this section, we analyze the designed microcavity structure using the theoretical methods outlined in the previous section. Our focus is on the entanglement properties of the quantum optical frequency comb generated by the cavity. As shown in Figure~\ref{fig:Q0}, we specifically consider the entanglement characteristics of the $\mathrm{TE_{00}}$ mode corresponding to ${L=1}$ in relation to the pump amplitude and pump frequency of the cavity, as depicted in Figure~\ref{fig:Q0}~(a). \textcolor{black}{According to Section~5.1, we mark the invalid region with grey areas where linearization may fail.} The \textcolor{black}{dark} green region represents areas with no entanglement, while regions with drastic color changes indicate modulation instability, where unstable entanglement components exist. Other colored regions denote areas where entanglement is tunable, with entanglement varying continuously with coordinates. Based on the characteristics of entanglement, we classify the system's states into three distinct phases: Non-Entanglement (NE) Phase, Modulation Instability (MI) Phase, and Entanglement-Tunable (ET) Phase.

In the ET and MI Phases, we select characteristic points and scan the other variable (pump amplitude or pump frequency) accordingly. Figure~\ref{fig:Q0}~(c, d, e, f) illustrate the relationship between pump amplitude or pump frequency and the intensity of the pump light inside the cavity. For points in the ET Phase, the scan lines do not pass through the MI Phase, and their curves change continuously without generating branches. However, for point in the MI Phase, both amplitude and frequency scans pass through the MI phase. As depicted in Figure~\ref{fig:Q0}~(e, f), multiple stable solutions below the threshold appear during the scan process.

Furthermore, we observe that entanglement is more easily detected at the boundaries between the three phases, while it tends to deteriorate closer to the boundaries with the original phases. Considering the robustness and tunability of the system, we aim to operate in the ET Phase to maximize the discovery of quantum entanglement in the process of generating quantum optical frequency combs.

\subsection{Single-Modal-Family Entanglement Control Consistency of QFCs}
\begin{figure}
  \includegraphics[width=\linewidth]{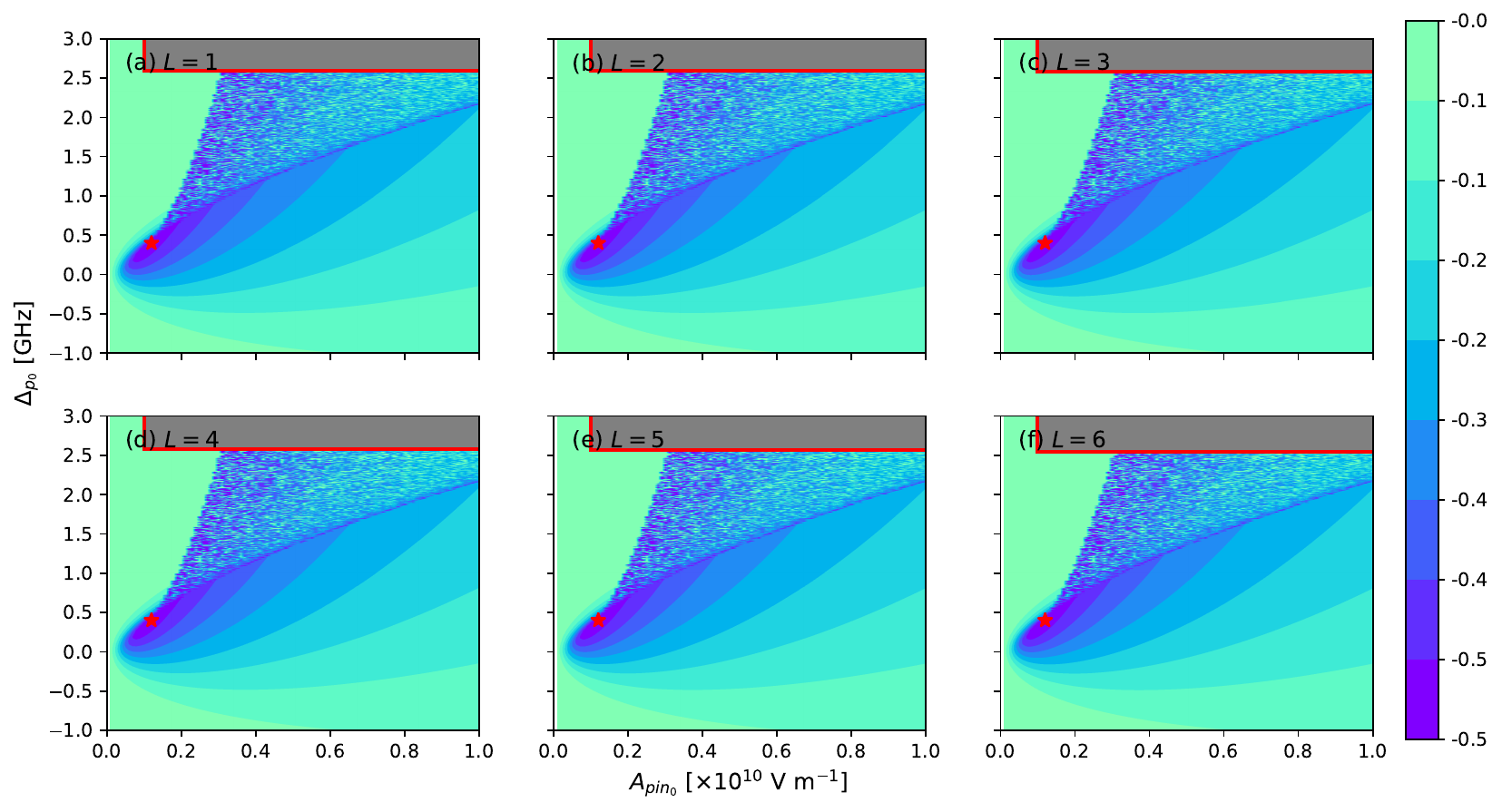}
  \caption{\label{fig:Q1} Heat map of the relationship between the quantum entanglement degree (${C_{min}}$), pump light frequency (${\Delta_{p_0}}$), and pump light amplitude (${A_{pin_0}}$) of the six pairs bipartite entangled comb teeth in the fundamental mode $\mathrm{TE_{00}}$, where color indicates the value of ${C_{min}}$, \textcolor{black}{gray region with red edges: invalid regions where linearization breaks down}. The star-marked points in all six figures are at the same position. (a, b, c, d, e, f) Generated six comb pairs near pump light, with label ${L=1,~2,~3,~4,~5,~6}$, follow 
almost the same distribution. The NE Phase area boarder as ${L}$ increase which is corresponding to theories that for the frequency far from the center of Lorentzian line shapes, it will be less sensitive to pump detuning.}
\end{figure}

In this section, we take the fundamental mode TE\(_{00}\) as an example to show how to maximize entanglement degrees. The relationship between \({C_{min}}\), \({\Delta_{p_0}}\), and \({A_{pin_0}}\) is crucial for understanding and controlling the bipartite entangled comb teeth in the fundamental mode TE\(_{00}\). Figure~\ref{fig:Q1} presents a comprehensive heat map illustrating this relationship across six pairs of bipartite entangled comb teeth for the fundamental mode TE\(_{00}\). Each subplot (a-f) corresponds to different values of \(L\) (ranging from 1 to 6), where \(L\) represents the label of the comb pairs.

In each subplot, the x-axis represents the pump amplitude \({A_{pin_0}}\) in units of \(  10^{10} \) V~m$^{-1}$, and the y-axis represents the pump detuning \(\Delta_{p_0} = \omega_{p_0} - \Omega_{p_0}\) in GHz. The color in the heat map indicates the value of \({C_{{min}}}\), with a color bar ranging from red (0.4) to purple (-0.5), as shown on the right side of the figure. This color gradient visually captures the sensitivity and the distribution of entanglement degree across different pump amplitudes and detuning. Unlike the heat map in Figure~\ref{fig:Q0}, the heat map in Figure~\ref{fig:Q1} is drawn using an interval method to better identify areas of strong entanglement.

The star-marked points in all six figures represent the same position, indicating the best pump conditions. The consistent pattern observed across the subplots suggests a similar distribution of the entanglement degree (\({C_{{min}}}\)) for different values of \({L}\). Specifically, the star-marked points denote the optimal conditions where the pump light parameters achieve a stable entanglement degree, essential for consistent QFCs.

Interestingly, as \({L}\) increases from 1 to 6, the ET Phase area broadens. This broadening is consistent with theoretical predictions that suggest the frequency regions far from the center of Lorentzian line shapes exhibit less sensitivity to pump detuning. This behavior highlights the robustness of the entanglement degree in maintaining consistency across varying pump conditions, crucial for practical implementations of QFCs in quantum communication and computation. The broader ET Phase area for higher \({L}\) values also suggests potential avenues for optimizing QFCs by strategically selecting pump conditions to achieve maximal entanglement efficiency.

\subsection{Simultaneously Maximizing Entanglement of QFCs in Distinct Modal Families}
\begin{figure}
  \includegraphics[width=\linewidth]{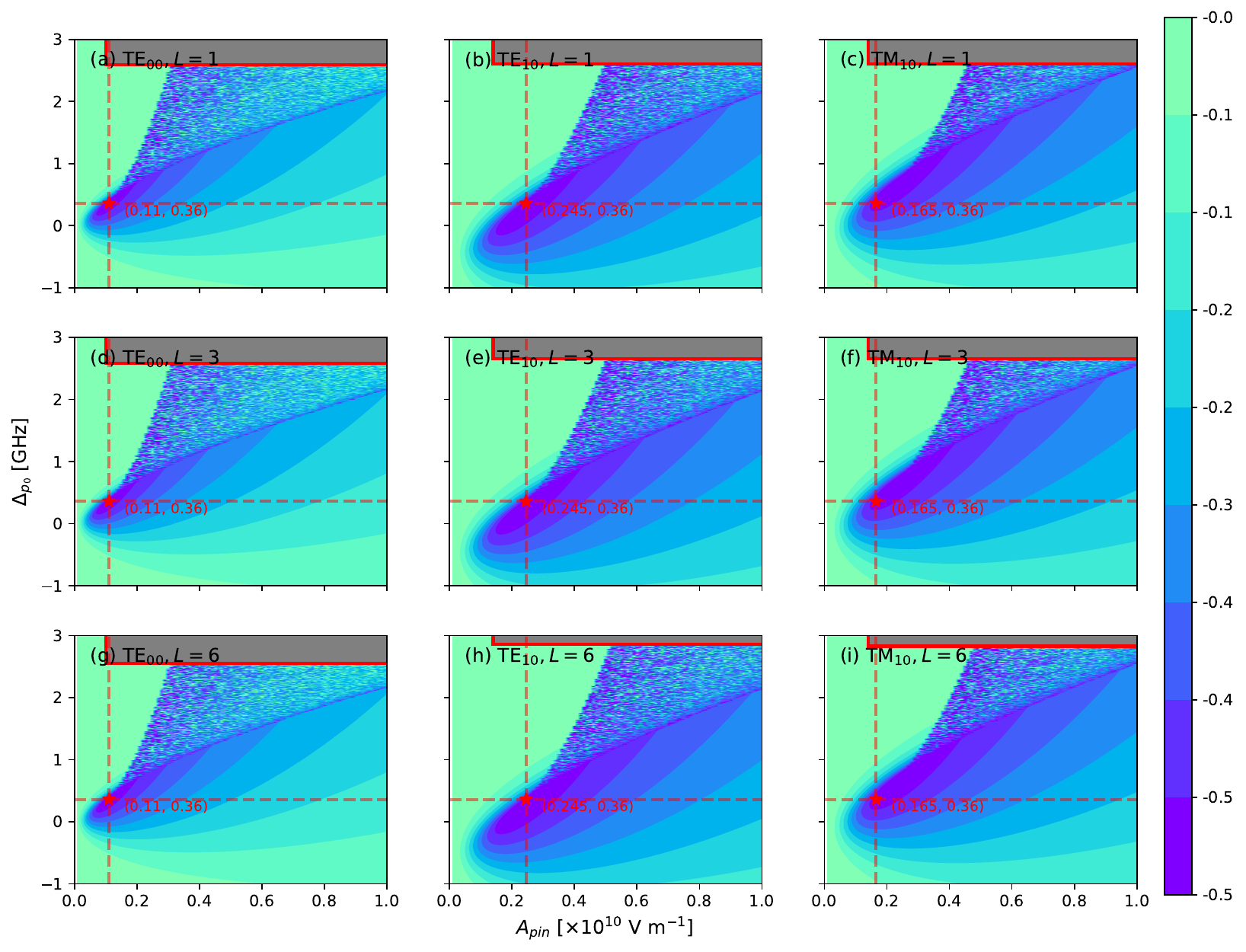}
  \caption{\label{fig:Q2}Heat map of the relationship between the quantum entanglement degree (${C_{min}}$), pump light frequency (${\Delta_{p_0}}$), and pump light amplitude (${A_{pin}}$) in different modal families, where color indicates the value of ${C_{min}}$, \textcolor{black}{gray region with red edges: invalid regions where linearization breaks down}. Here we choose only three pairs comb teeth of each modal family, considering single-modal-family entanglement control
consistency. Because pump frequency of the three modal families must be the same, the only entanglement tuning method is tuning pump light amplitude when the frequency fixed. The red star mark in each figure gives out the maximal entanglement degree in the ET Phase, as we avoid near the MI Phase for robustness. The best pump condition for modal families is $\text{``}{\Delta_{p_0}=0.36~\mathrm{GHz}}$, $\mathrm{TE_{00}}:A_{pin}=1.1\times 10^{9}~\mathrm{V~m^{-1}}$, $\mathrm{TE_{10}}:A_{pin}=2.45\times 10^{9}~\mathrm{V~m^{-1}}$, $\mathrm{TM_{10}}:A_{pin}=1.65\times 10^{9}~\mathrm{V~m^{-1}}\text{''}$.}
\end{figure}

\begin{figure}
  \includegraphics[width=\linewidth]{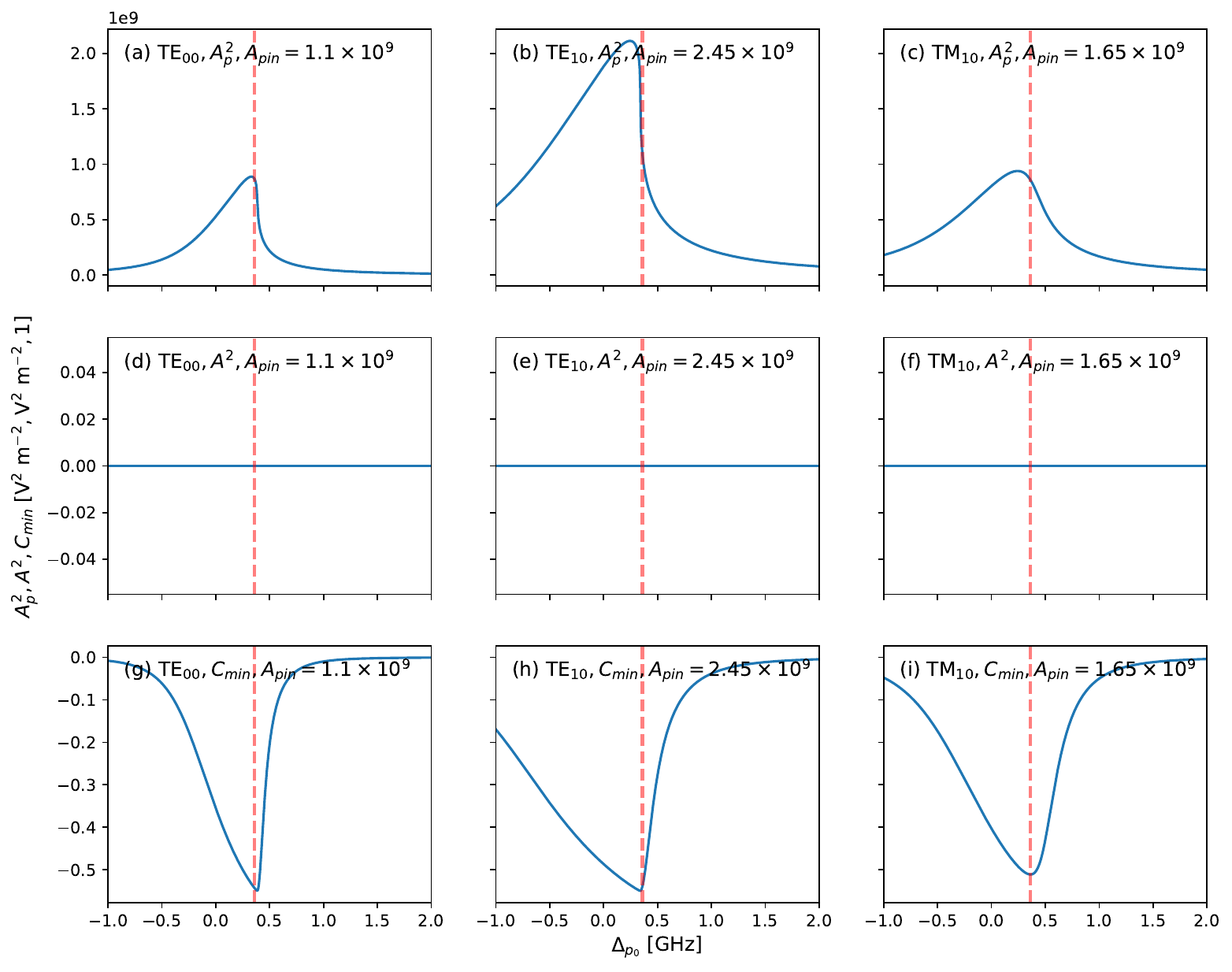}
  \caption{\label{fig:Q3}Dynamics of pump and signal/idler light power ${A_p^2}$ and $\mathrm{A^2}$, and entanglement degree at the best pump amplitude condition and ${L=1}$. (a, b, c) Pump light power in the cavity ${A_p^2}$. When tuning pump frequency from the ET Phase to the NE Phase, the power first increases and then decreases, and no bi-stability found here. (d, e, f) Signal/Idler light power in the cavity, all tuning processes are under threshold. (g, h, i) All modal families arrive maximal entanglement at our best pump frequency, when ${A_p^2}$ is not the maximum value.}
\end{figure}

In this section, we delve into the challenge of simultaneously maximizing entanglement in QFCs across distinct modal families. Figures~\ref{fig:Q2} and \ref{fig:Q3} provide comprehensive visualizations and analysis to illustrate the intricate dynamics involved in achieving optimal entanglement conditions.

Figure~\ref{fig:Q2} presents a heat map showcasing the relationship between \({C_{min}}\), \({\Delta_{{p_0}}}\), and \({A_{{pin}}}\) across different modal families. Each subplot represents a combination of modal families (TE\(_{00}\), TE\(_{10}\), TM\(_{10}\)) with varying \(L\) values ($L=1,~3,~6$). The color gradient indicates the value of \({C_{min}}\), with red marking the relatively high entanglement degree.

In each subplot, the x-axis denotes the pump amplitude \(A_{{pin}}\) in units of \( \times 10^{10} \) V~m$^{-1}$, and the y-axis denotes the pump detuning \(\Delta_{{p_0}} = \omega_{{p_0}} - \Omega_{{p_0}}\) in GHz. As we use a monochromatic pump light to excite all modal families, when we optimize entanglement \({\Delta_{{p_0}}}\) must be the same. The red dashed lines and the star markers indicate the optimal conditions for maximizing entanglement. The figures reveal that the best pump conditions for the modal families are \(\Delta_{{p_0}} = 0.36\) GHz, \(A_{{pin}} = 1.1 \times 10^9\) V~m$^{-1}$ for TE\(_\mathrm{00}\), \(A_{{pin}} = 2.45 \times 10^9\) V~m$^{-1}$ for TE\(_\mathrm{10}\), and \(A_{{pin}} = 1.65 \times 10^9\) V~m$^{-1}$ for TM\(_\mathrm{10}\). This optimal tuning ensures maximal entanglement degrees while avoiding the MI Phase for robustness.

The detailed analysis of Figure~\ref{fig:Q2} also indicates that for higher \(L\) values, the ET Phase area broadens. The reason for this broadening is the same as explained in the previous section.

Figure~\ref{fig:Q3} further explores the dynamics of pump and comb teeth pairs power \({A_{{p}}^2}\) and \({A^2}\), along with the entanglement degree at the best pump amplitude conditions for \(L = 1\) of all three modal families. The subplots illustrate the variations in cavity dynamics for TE\(_{00}\), TE\(_{10}\), and TM\(_{10}\) modal families. The top row (a-c) shows the pump light power in the cavity \(A_{{p}}^2\) as a function of pump detuning with pump amplitude $A_{pin}$ fixed. The middle row (d-f) represents the comb teeth pairs power in the cavity, highlighting that all tuning processes are below the threshold and avoiding bi-stability. The bottom row (g-i) shows the entanglement degree \({C_{min}}\), where each modal family achieves maximal entanglement at the identified optimal pump conditions.

Specifically, Figure~\ref{fig:Q3} reveals that when tuning the pump frequency from the ET Phase to the NE Phase, the power in the cavity initially increases and then decreases, with no bi-stability observed. Then we delve into strategies aimed at maximizing entanglement within QFCs across distinct modal families within the microcavity structure. While the previous discussion focused on entanglement control within a single modal family, here we extend our analysis to encompass multiple modal families. By maximizing entanglement across these distinct families, we aim to enhance the versatility and functionality of QFCs for quantum applications.

\section{Discussion}
In our setup, we carefully choose the resonance peaks of our pump, ensuring that the peaks of the three distinct modal families overlap effectively at the pump wavelength. Since our simulations are based on a near-cold cavity scenario, they do not account much for SPM, XPM and Bragg Scattering effects that typically arise when operating at high power levels. This approach simplifies our analysis and allows us to evaluate the system's behavior in the low-power regime without interference from Kerr nonlinear phenomena. However, high pump power is also considerable because we have included SPM and XPM as an add-on in the Hamiltonian for determining threshold, as shown in Equation~(\ref{eq:hamil}). As a result, we can use absolute detuning to cold pump resonance peak, without using relative detuning to hot moving peak. 

At low pump power, due to the effective overlap of modal families' resonances at the pump wavelength, we can use monochromatic pump light to excite them. Consequently, our system only needs an additional SLM to modulate the pump profile, allowing a regular microcavity to generate spatially multiplexed quantum entangled optical frequency combs. Moreover, our simulation is based on steady-state equations, which indicate the stable states the system will be in under thermal equilibrium. In the NE and ET Phase, there is only one solution below the threshold, but in the MI Phase, there are multiple solutions, all of which are below the threshold.

The influence of fabrication errors on the performance of WGMR systems must also be carefully considered. One of the most significant impacts of fabrication errors is on the geometric dimensions of the resonator, such as variations in the radius and cross-sectional shape. These deviations can alter the designed dispersion, resulting in shifts in the resonance frequencies. A key objective of this work is to achieve the overlap of as many resonant peaks from different modal families as possible. However, such fabrication errors could severely affect this, as resonance shifts directly influence whether a single monochromatic pump can efficiently excite multiple spatial modes simultaneously.

Additionally, the matching between the shapes of the resonator and the waveguide significantly affects the coupling efficiency between modes and the overall Q-factor, further impacting the spatial mode distribution of the pump. Variations in the coupling due to geometry changes could affect the balance of pump spatial modes. Moreover, increased surface roughness will lead to a reduction in the Q-factor, which in turn affects the efficiency and stability of frequency comb generation. Higher-order mode coupling is particularly sensitive to fabrication errors, which may lead to notable changes in the multi-mode coupling characteristics.

To address these issues, several potential compensation and optimization strategies can be employed. First, adjusting the pump wavelength can partially compensate for the frequency shifts and variations in mode coupling efficiency caused by geometric errors. Second, utilizing high-precision fabrication techniques can minimize the impact of surface roughness on the Q-factor. Furthermore, in experimental settings, fabrication-induced errors can be mitigated through precise calibration of the system’s transmission spectra. By measuring the transmission spectra of the microring resonator, the actual coupling rates and loss rates of different modes can be fitted, allowing for more accurate simulations that replace the original design parameters. Finally, by using a SLM to modulate the pump light, we provide additional flexibility for optimizing the spatial mode distribution of the pump, offering greater control in subsequent pump mode adjustments.

\section{Conclusion}
We build an on-chip $\mathrm{Si_3 N_4}$ WGMR capable of supporting multiple modal families. Quantum entangled frequency combs can be formed around the pump frequency based on the third-order nonlinearity. With continuous-wave and below-threshold pumping, in each modal family, we can generate bipartite-type entangled quantum frequency combs which are useful in both discrete-variable and continuous-variable quantum processing. We employ the bipartite entanglement criterion to quantify the entanglement of bipartite comb teeth in each modal family. In our scheme, we do not need to consider nonlinear interactions across distinct modal families, as the pump is monochromatic. As shown in our simulation and analysis, there is no complex entanglement structure beyond the bipartite type within below-threshold pumping. With our quantum frequency comb, we are able to provide quantum entangled states in no fewer than twelve channels of each modal family, and in our design, we support three modal families as multiplex channels in a single WGMR, making it a suitable solution for multi-channel quantum information networks. Our approach makes use of higher-order modal families to generate quantum frequency combs which is not used in traditional quantum frequency comb generation methods and is capable of generating high-density entanglement. Among these channels, we can optimize the entanglement degree of each mode at any stage under certain initially set injected pump power and pump detuning through temperature adjustment, which may inspire better quantum resources and lead to better understanding of the entanglement mechanism.




\medskip
\textbf{Acknowledgements} \par 

This work is supported by the Key-Area Research and Development Program of Guangdong Province
(Grant No.2018B030325002), the National Natural Science Foundation of China (Grant No.62075129,
61975119), the Open Project Program of SJTU-Pinghu Institute of Intelligent Optoelectronics (Grant
No.2022SPIOE204) and the Sichuan Provincial Key Laboratory of Microwave Photonics (Grant 2023-04), and the Science and Technology on Metrology and Calibration Laboratory (Grant No. JLKG2024001B002).

\medskip
\textbf{Conflict of Interest} \par 
The authors declare no conﬂict of interest.

\medskip
\textbf{Data Availability Statement} \par 
The data that support the ﬁndings of this study are available from the corresponding author upon reasonable request.

\medskip

%
\bibliographystyle{MSP}
\bibliography{main}



\newpage

\begin{table}
 \caption{\label{tab:tableC1}%
Cavity structure parameters shown in Figure~\ref{fig:C0}~(a)~\ding{193}. ${R}$ is the radius of the whole ring structure.}
  \begin{tabular}[htbp]{@{}lll@{}}
    \hline
    \textrm{Cavity structure parameters}&
\textrm{Values}& 
\textrm{Units}\\
    \hline
    ${W_w}$ & $\mathrm{0.8}$ & $\mathrm{\mu m}$\\
${W_h}$ & $\mathrm{1.8}$ & $\mathrm{\mu m}$\\
${\theta}$ & $\mathrm{89}$ & $\mathrm{ deg}$\\
${C_b}$ & $\mathrm{1.7}$ & $\mathrm{\mu m}$\\
${C_h}$ & $\mathrm{6.5}$ & $\mathrm{\mu m}$\\
${C_w}$ & $\mathrm{8.0}$ & $\mathrm{\mu m}$\\
${R}$ & $\mathrm{240}$ & $\mathrm{\mu m}$\\
${d}$ & $\mathrm{490}$ & $\mathrm{n m}$\\
    \hline
  \end{tabular}
\end{table}

\begin{table}
 \caption{\label{tab:tableC2}%
Cavity parameters generated by $COMSOL ~Multiphysics$ simulation for all supported modal families.}
  \begin{tabular}[htbp]{@{}lllll@{}}
    \hline
    Parameters&
$\mathrm{Mode~1~(TE_{00})}$& 
$\mathrm{Mode~2~(TM_{00})}$& 
$\mathrm{Mode~3~(TE_{10})}$& 
$\mathrm{Mode~4~(TM_{10})}$\\
    \hline
    ${D_1}~\mathrm{[rad~s^{-1}]}$ & $\mathrm{6.02 \times 10^{11}}$ & $\mathrm{5.94 \times 10^{11}}$ & $\mathrm{5.79 \times 10^{11}}$ & $\mathrm{5.76 \times 10^{11}}$\\
${D_2}~\mathrm{[rad~s^{-1}]}$ & $\mathrm{2.57 \times 10^6}$ & $\mathrm{5.64 \times 10^6}$ & $\mathrm{1.48 \times 10^7}$ & $\mathrm{1.28 \times 10^7}$\\
${D_3}~\mathrm{[rad~s^{-1}]}$ & $\mathrm{-4.34 \times 10^3}$ & $\mathrm{5.04 \times 10^2}$ & $\mathrm{-2.43 \times 10^3}$ & $\mathrm{5.95 \times 10^3}$\\
${D_4}~\mathrm{[rad~s^{-1}]}$ & $\mathrm{-7.45 \times 10^0}$ & $\mathrm{-3.32 \times 10^1}$ & $\mathrm{-5.69 \times 10^1}$ & $\mathrm{-7.72 \times 10^1}$\\
${D_5}~\mathrm{[rad~s^{-1}]}$ & $\mathrm{3.20 \times 10^{-2}}$ & $\mathrm{1.22 \times 10^{-1}}$ & $\mathrm{3.06 \times 10^{-1}}$ & $\mathrm{3.28 \times 10^{-1}}$\\
$fsr~\mathrm{[GHz]}$ & $\mathrm{9.58 \times 10^{1}}$ & $\mathrm{9.45 \times 10^{1}}$ & $\mathrm{9.21 \times 10^{1}}$ & $\mathrm{9.17 \times 10^{1}}$\\
${f_0}~\mathrm{[THz]}$ & $\mathrm{2.145933 \times 10^{2}}$ & $\mathrm{2.146082 \times 10^{2}}$ & $\mathrm{2.145928 \times 10^{2}}$ & $\mathrm{2.145932 \times 10^{2}}$\\
${\lambda_0}~\mathrm{[nm]}$ & $\mathrm{1.397026 \times 10^{3}}$ & $\mathrm{1.396929 \times 10^{3}}$ & $\mathrm{1.397029 \times 10^{3}}$ & $\mathrm{1.397027 \times 10^{3}}$\\
${A_{eff}}~\mathrm{[\mu m^2]}$ & $\mathrm{1.13 \times 10^{0}}$ & $\mathrm{1.30 \times 10^{0}}$ & $\mathrm{1.38 \times 10^{0}}$ & $\mathrm{1.37 \times 10^{0}}$\\
${n_{eff}}~\mathrm{[\mu m^2]}$ & $\mathrm{1.86 \times 10^{0}}$ & $\mathrm{1.84 \times 10^{0}}$ & $\mathrm{1.77 \times 10^{0}}$ & $\mathrm{1.75 \times 10^{0}}$\\
${\eta}~\mathrm{[m~W^{-1}]}$ & $\mathrm{9.59 \times 10^{-1}}$ & $\mathrm{8.28 \times 10^{-1}}$ & $\mathrm{7.83 \times 10^{-1}}$ & $\mathrm{7.86 \times 10^{-1}}$\\
${g_0}~\mathrm{[\mu m^2]}$ & $\mathrm{2.34 \times 10^0}$ & $\mathrm{2.08 \times 10^0}$ & $\mathrm{2.13 \times 10^0}$ & $\mathrm{2.17 \times 10^0}$\\
${Q}$ & $\mathrm{1.00 \times 10^6}$ & $\mathrm{1.00 \times 10^6}$ & $\mathrm{5.00 \times 10^5}$ & $\mathrm{5.00 \times 10^5}$\\
${\mu/(\gamma+\mu)}$ & $\mathrm{4.50 \times 10^{-1}}$ & $\mathrm{4.50 \times 10^{-1}}$ & $\mathrm{4.50 \times 10^{-1}}$ & $\mathrm{4.50 \times 10^{-1}}$\\
    \hline
  \end{tabular}
\end{table}





\begin{figure}
\textbf{Table of Contents}\\
\medskip
  \includegraphics{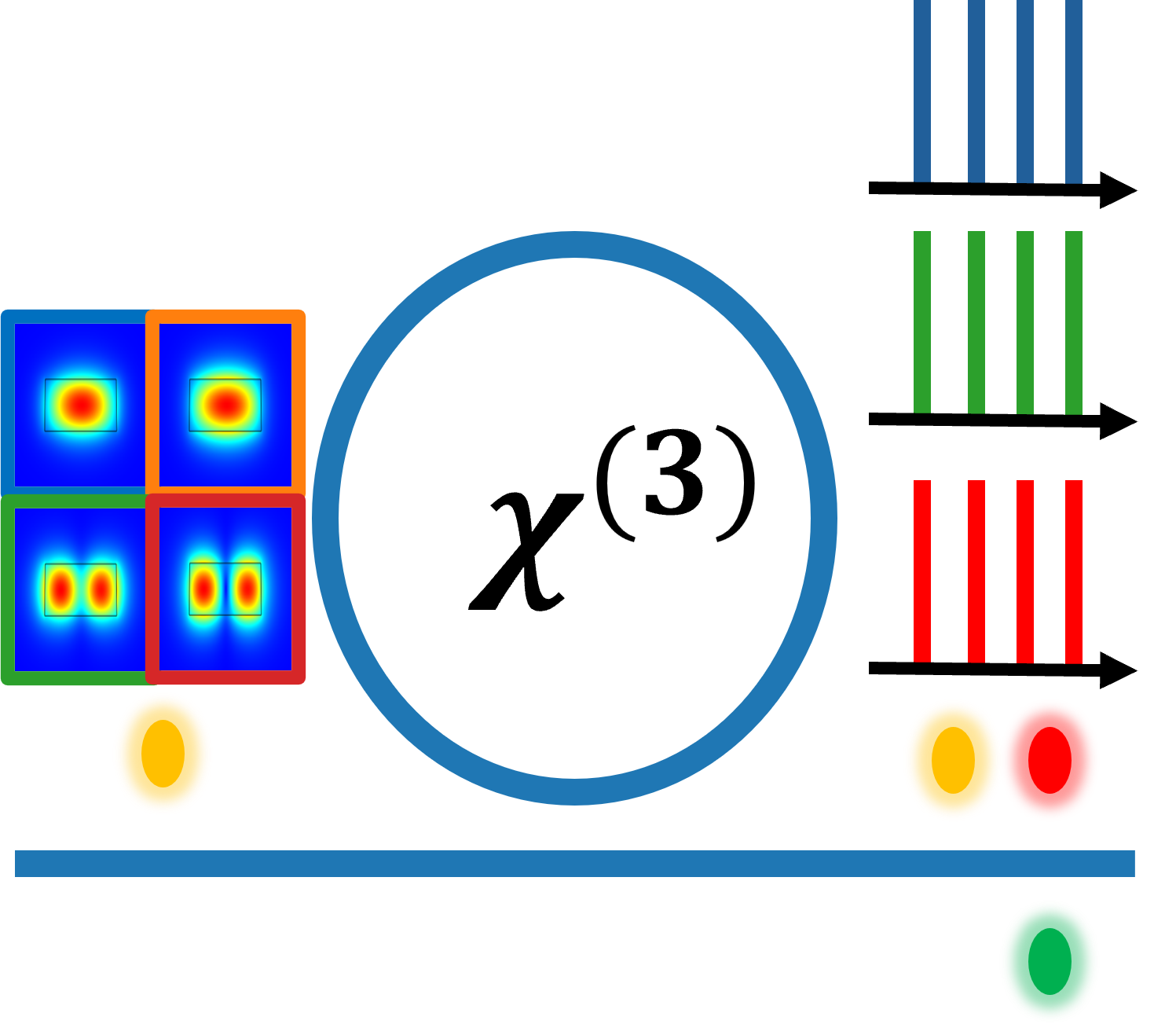}
  \medskip
  \caption*{The research presents the generation of quantum entangled frequency combs in a silicon nitride WGMR. Under profile-modulated monochromatic, below-threshold continuous-wave pumping, bipartite entanglement forms across multiple modal families, supporting twelve channels per family. This approach enhances multi-channel quantum network capabilities and allows optimization of the entanglement degree through pump intensity and frequency adjustments, providing better quantum resources.}
\end{figure}

\end{document}